\documentclass[11pt,letterpaper]{article}
\usepackage{amsmath,amssymb,graphicx,fullpage,multirow,authblk}
\usepackage[round]{natbib}

\begin{document}
\title{Unsaturated Hydraulic Conductivity Models Based on Truncated Lognormal Pore-size Distributions}
\author{Bwalya Malama%
\thanks{\texttt{bnmalam@sandia.gov}}} 
\affil{Performance Assessment Department, Sandia National Laboratories, Carlsbad NM, USA}
\author{Kristopher L. Kuhlman%
\thanks{\texttt{klkuhlm@sandia.gov}}}
\affil{Applied Systems Analysis \& Research Department, Sandia National Laboratories, Albuquerque NM, USA}
\maketitle

\begin{abstract}
We develop a closed-form three-parameter model for unsaturated hydraulic conductivity associated with the Kosugi three-parameter lognormal moisture retention model. The model derivation uses a slight modification to Mualem's theory, which is nearly exact for non-clay soils. Kosugi's three-parameter lognormal moisture retention model uses physically meaningful parameters, but a corresponding closed-form relative hydraulic conductivity model has never been developed. The model is further extended to a four -parameter model by truncating the underlying pore size distribution at physically permissible minimum and maximum pore radii. The proposed closed-form models are fitted to well-known experimental data, to illustrate their utility. They have the same physical basis as Kosugi's two-parameter model, but are more general.
\end{abstract}

\section{Introduction}
Understanding and predicting infiltration of water into unsaturated soil is critical to both agricultural and groundwater-hydrology applications. Precipitation infiltrates through the vadose zone, often carrying contaminants from the surface to regional aquifers. Numerically simulating moisture redistribution in the vadose zone using Richards' equation requires functions relating soil water content and relative hydraulic conductivity to capillary pressure head (e.g., \citet[\S2.5]{warrick2003}). \cite{durner2006} summarize techniques to collect data characterizing these unsaturated flow relationships in soils. Many different functional forms have been proposed to capture the essential characteristics of observed soil behavior, allowing for efficient and accurate predictive simulation. Although any arbitrary function can be adopted to represent a soil's behavior during infiltration, simpler mathematical forms often allow closed-form representation of the moisture retention and relative hydraulic conductivity curves. Common closed-form functions include models by \cite{gardner1958}, \cite{brooks64}, \cite{vangenuchten1980}, and \cite{kosugi1994}. Closed-form expressions for unsaturated hydraulic conductivity allow straightforward implementation in numerical models, avoiding costly and error-prone numerical integration. Closed-form expressions can provide more insight into the relationship between parameters in the moisture retention and hydraulic conductivity models than purely numerical schemes (e.g., \cite{priesack2006}).

\citet{kosugi1994} assumed soil pore size is a lognormal random variable and derived a physically based three-parameter model for moisture retention, the three parameters being the mean and variance of the pore-size distribution and the maximum pore radius. In the limiting case where the maximum pore radius becomes infinite the three-parameter model simplifies to a two-parameter model for which \citet{kosugi1996} developed the closed-form expression for unsaturated hydraulic conductivity using the theory of \citet{mualem1976}. \citet{kosugi1996} did not develop a three-parameter closed-form equation for hydraulic conductivity, but reverted to the two-parameter form, owing to difficulty in analytically integrating the expression of \citet{mualem1976}.  We extend the work of \citet{kosugi1994,kosugi1996}, developing closed-form expressions for unsaturated hydraulic conductivity associated with the three-parameter lognormal moisture retention model. The derivation of the closed-form equation for unsaturated hydraulic conductivity is made possible by an approximation to the theory of \citet{mualem1976}. Further, we modify the pore-size probability density function (PDF) of \citet{kosugi1994} by incorporating a nonzero minimum pore radius, as suggested by \citet{brutsaert1966}. This modification results in a four-parameter moisture retention model and a corresponding four-parameter closed-form equation for unsaturated hydraulic conductivity, again obtained using the modified theory of \citet{mualem1976}.  The four parameters are all based on porous medium properties like \citet{kosugi1996}; they are not fitting parameters without physical significance.  The four-parameter model is a generalization of the two-parameter model of \citet{kosugi1996} and simplifies to it when the minimum and maximum pore sizes tend to zero and infinity.

\section{Theory}
The lognormal distribution is commonly used to statistically characterize pore size in granular porous media. \citet{brutsaert1966} and \citet{kosugi1994,kosugi1996} considered  lower- and upper-tail truncated lognormal PDFs for pore-size distributions. \citet{brutsaert1966} considered the log-transformed random pore radius $R-r_0$, where $r_0$ is the radius at which the effective moisture content vanishes (associated with residual saturation). 

\subsection{Three-parameter lognormal model}
The classical (non-truncated) lognormal pore-size distribution  \citep{brutsaert1966} is measured here by the random pore radius $R\in [0,\infty]$. In a physically realistic porous medium $R\in [0,r_\mathrm{max}]$, where $r_\mathrm{max}$ is some finite maximum pore radius. To account for the finite interval, \citet{kosugi1994} used the random variable $R_e = \left(1/R - 1/r_\mathrm{max}\right)^{-1}$ to rescale the classical PDF.  The PDF of $R$ is related to the PDF of $R_e$ as
\begin{equation}
f_R(r)=\frac{f_{R_e}\left[\left(1/r-1/r_\mathrm{max}\right)^{-1}\right]}{\left(1-r/r_\mathrm{max}\right)^2}.
\end{equation}
According to Young-Laplace theory, capillary pressure head $h$ and pore radius $r$ are related by $h = \kappa / r$, where $\kappa=2\gamma \cos \alpha/(\rho g)$, $\gamma$ is interface surface tension, $\alpha$ is the interface contact angle, $\rho$ is fluid density, and $g$ is gravitational acceleration.  For water in a glass tube, $\kappa\approx0.149\,\mathrm{cm}^2$. It follows the PDF of the random capillary pressure head $H$ is
\begin{equation}
f_H(h)=\frac{f_{R_e}\left[\left(h/\kappa - 1/r_\mathrm{max}\right)^{-1}\right]}{\left(h/\kappa-1/r_\mathrm{max}\right)^2}.
\end{equation}
Further, if $R_e$ is lognormally distributed, the PDF for the capillary pressure head $H$ can be written as
\begin{equation}
\label{eqn:lognH_proposed}
f_H(h) = \frac{1}{\sqrt{2\pi} \sigma_Z (h-h_c)}\exp \left[-\left(\frac{\ln\left(h-h_c\right) - \mu_\eta}{\sqrt{2}\sigma_Z}\right)^2\right],
\end{equation}
for all $h>h_c$, where $h_c = \kappa/r_\mathrm{max}$ is the bubbling pressure head, $\mu_\eta=\ln(\kappa) - \mu_Z$ is the mean of $\ln(H)$, $\sigma_Z^2$ is the variance of $Z$, $\mu_Z$ is the mean of $Z$, and $Z=\ln\left( R_e\right)$. \citet{kosugi1994} used the dimensionless random variable $R'_e=R_e/r_\mathrm{max}$ to obtain the PDF in (\ref{eqn:lognH_proposed}) with the parameters $\mu_Z$ and $\sigma_Z^2$ scaled appropriately.

\citet{kosugi1994} showed the three-parameter moisture retention curve related to (\ref{eqn:lognH_proposed}) is
\begin{equation}
\label{eqn:proposed_mrc1}
\theta^{\ast}(h) = \left\{
\begin{array}{ll}
\frac{1}{2}\mathrm{erfc}\left(\frac{\ln(h-h_c)-\mu_\eta}{\sqrt{2} \sigma_Z}\right)  & \; h>h_c ,\\
1 &\; h \leq h_c ,
\end{array} \right. 
\end{equation}
where $\theta^{\ast}(h)=(\theta(h)-\theta_r)/(\theta_s - \theta_r)$ is moisture capacity, $\theta(h)$ is volumetric moisture content, $\theta_r$ is residual moisture content, $\theta_s$ is saturated moisture content, and erfc is the complementary error function (e.g., \citet[\S7]{abramowitz64}).  \citet{kosugi1994} did not develop a corresponding closed-form equation for unsaturated hydraulic conductivity from \ref{eqn:lognH_proposed}.

\citet{mualem1976} developed a widely used functional relation between unsaturated hydraulic conductivity $K(\theta^{\ast})=K_s K_r$ and $h$ 
\begin{equation}
\label{eqn:mualem}
K_r(\theta^{\ast})=\sqrt{\theta^{\ast}}\left[\left(\int_0^{\theta^{\ast}}\frac{\mathrm{d}x}{h(x)}\right) \bigg / \left(\int_0^1 \frac{\mathrm{d}x}{h(x)}\right) \right]^2,
\end{equation}
where $K_r$ and $K_s$ are relative and  saturated hydraulic conductivity and $x$ is an integration variable. Equation (\ref{eqn:mualem}) can be rewritten in terms of the capillary pressure head PDF as
\begin{equation}
\label{eqn:mualem2}
K_r(\theta^{\ast})=\sqrt{\theta^{\ast}}\left[\left(\int_h^\infty \frac{f_H(x)}{x} \mathrm{d}x\right) \bigg / \left(\int_0^\infty \frac{f_H(x)}{x}\mathrm{d}x \right) \right]^2.
\end{equation}
Using the theory of \citet{mualem1976}, \citet{kosugi1996} developed the two-parameter closed-form equation for unsaturated hydraulic conductivity by setting $r_\mathrm{max} \rightarrow \infty$. \citet{kosugi1996} made this simplification because the theory of \citet{mualem1976} as given in (\ref{eqn:mualem2}) is not readily amenable to integration when $r_\mathrm{max}$ is finite.

\begin{figure}[h]
\includegraphics[width=0.5\textwidth]{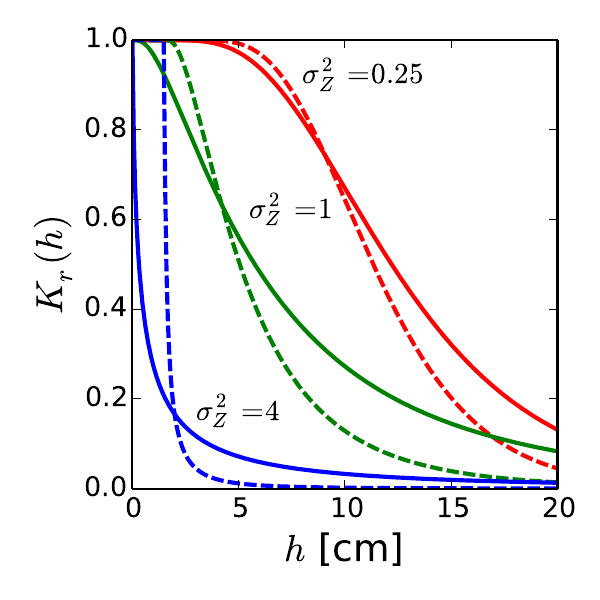}
  \caption{\label{fig:kosugi_K} Predicted $K_r(h)$ for the two-parameter model of \citet{kosugi1996} (solid), and the proposed three-parameter lognormal model for finite $r_\mathrm{max}$ (dashed).}
\end{figure}

We obtain an approximate closed-form expression for unsaturated hydraulic conductivity for the lognormal PDF with finite values of $r_\mathrm{max}$ by modifying (\ref{eqn:mualem}) due to \citet{mualem1976} into 
\begin{equation}
\label{eqn:mualem3}
K_r(\theta^{\ast})=\sqrt{\theta^{\ast}}\left[\left(\int_0^{\theta^{\ast}}\frac{\mathrm{d}x}{h(x)-h_c}\right) \bigg/ \left(\int_0^1 \frac{\mathrm{d}x}{h(x)-h_c}\right) \right]^2,
\end{equation}
based on the assumption $f_H(h)/h \approx f_H(h)/(h-h_c)$.  Using this approximation the relative hydraulic conductivity for the three-parameter lognormal model is
\begin{equation}
\label{eqn:lognK_kosugi}
K_r(h) \backsimeq \left \{
\begin{array}{ll}
 \sqrt{\theta^{\ast}}\left\lbrace\frac{1}{2}\mathrm{erfc} \left[\frac{\ln(h-h_c)-\mu_\eta + \sigma_Z^2 }{\sqrt{2} \sigma_Z }\right]\right\rbrace^2 &\; h>h_c ,\\
1 &\; h \leq h_c .
\end{array} \right. 
\end{equation} 
In the limit as $r_\mathrm{max} \rightarrow \infty$, $h_c \rightarrow 0$, the assumption given above is exact, and (\ref{eqn:lognK_kosugi}) reduces to the two-parameter unsaturated hydraulic conductivity expression derived by \citet{kosugi1996}. The approximation introducted by the assumption leading to (\ref{eqn:mualem3}) is best when $r_\mathrm{max}$ is relatively large ($h_c$ is small). Figure \ref{fig:kosugi_K} shows that the truncated lognormal pore-size distribution (\ref{eqn:lognK_kosugi}) shifts the $K_r$ curves ($K_r=1$ at $h=h_c$ -- dashed curves) compared to the solid curves corresponding to the two-parameter model of \citet{kosugi1996}, where $K_r=1$ is reached at $h=0$.

\subsection{Four-parameter lognormal model}
\begin{figure}[h]
\includegraphics[width=0.5\textwidth]{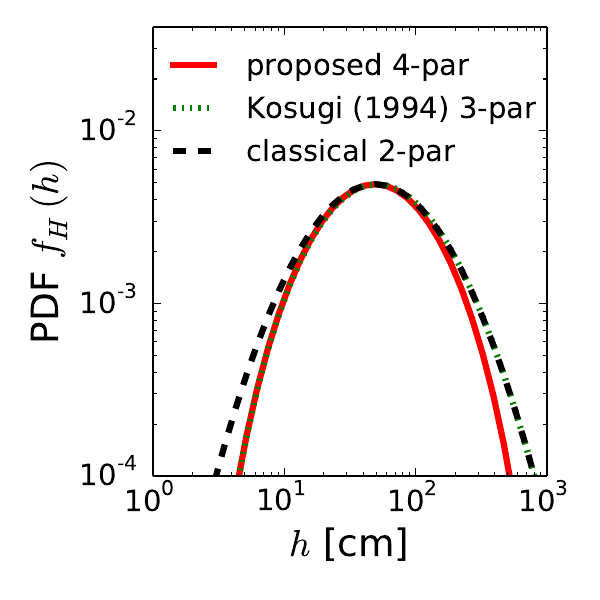}  
  \caption{\label{fig:logn_densityz} Comparison of lognormal capillary pressure head PDFs for the classical, three-parameter \citep{kosugi1994}, and proposed four-parameter (\ref{eqn:lognH_proposed2}) models.}
\end{figure}

We incorporate the lower-tail truncation of \citet{brutsaert1966} into the distribution of \citet{kosugi1994}, by introducing the random variable 
\begin{equation}
\hat{R}_e = \left(\frac{1}{R-r_0}-\frac{1}{r_\mathrm{max}}\right)^{-1},
\end{equation}
which yields the following lognormal PDF for capillary pressure head,
\begin{equation}
\label{eqn:lognH_proposed2}
f_H(h) = \frac{1}{\sqrt{2\pi} u \sigma_Z}\exp \left[-\left(\frac{\ln(u) - \mu_\eta}{\sqrt{2}\sigma_Z}\right)^2\right],
\end{equation}
for all $h \in [h_c, h_\mathrm{max}]$ where  $u=(1/h-1/h_\mathrm{max})^{-1}-h_c$ and $h_\mathrm{max}=\kappa /r_0$ is the pressure head associated with the smallest undrainable pores. Figure \ref{fig:logn_densityz} shows the three lognormal PDFs for capillary pressure head: the classical (non-truncated) lognormal distribution, the upper-truncated lognormal distribution (\ref{eqn:lognH_proposed}), and the doubly truncated lognormal distribution (\ref{eqn:lognH_proposed2}). It can be seen the three-parameter model of \citet{kosugi1994} departs from the classical lognormal distribution only at small head values (large pore radii) whereas the proposed four-parameter distribution departs from the classical distribution at both the lower and upper limbs of the function.  The four-parameter model can be considered most physically realistic, while the two- and three-parameter models are simplifications.  For some soils the simplified PDFs may be adequate.

A moisture retention model is derived from (\ref{eqn:lognH_proposed2}) in a similar manner, and is given by
\begin{equation}
\label{eqn:proposed_mrc2}
\theta^{\ast}(h) = \left \{ 
\begin{array}{ll}
\frac{1}{2}\mathrm{erfc}\left[\frac{\ln(u)-\mu_\eta}{\sqrt{2} \sigma_Z}\right] &\;  h_c < h < h_\mathrm{max},\\
1 &\; h \le h_c,\\
0 &\; h \ge h_\mathrm{max}.
\end{array} \right.
\end{equation}
Finally, it can be shown that the closed-form expression for unsaturated hydraulic conductivity using the doubly truncated PDF (\ref{eqn:lognH_proposed2}) and the modified equation of \citet{mualem1976} (\ref{eqn:mualem2}) is
\begin{equation}
\label{eqn:lognK_proposed}
K_r(h) \backsimeq  \left \{
\begin{array}{ll}
\sqrt{\theta^{\ast}}\left\lbrace\frac{1}{2}\mathrm{erfc} \left[\frac{\ln(u)-\mu_\eta - \sigma_Z^2}{\sqrt{2} \sigma_Z }\right]\right\rbrace^2  & \; h_c < h < h_\mathrm{max},\\
1 & \; h \le h_c, \\
0 & \; h \ge h_\mathrm{max}.
\end{array} \right.
\end{equation}
In the limit as both $r_\mathrm{max}\rightarrow \infty$ and $r_0 \rightarrow 0$, (\ref{eqn:proposed_mrc2}) and (\ref{eqn:lognK_proposed}) simplify to corresponding two-parameter expressions from \citet{kosugi1994,kosugi1996}.

\section{Fitting lognormal models to experimental data}
\begin{figure}[h]
\includegraphics[width=\textwidth]{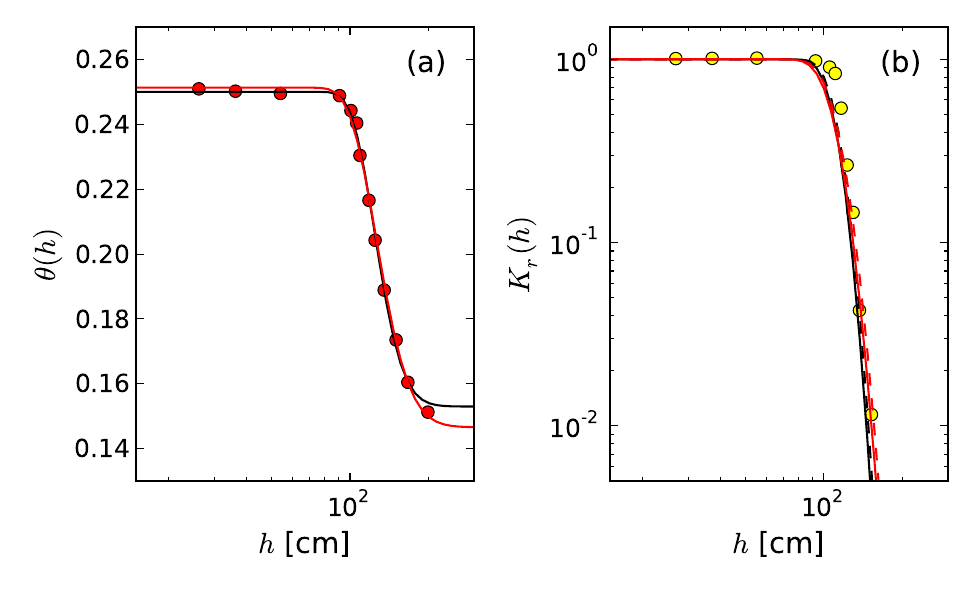}
  \caption{\label{fig:hygsanstone} Hygiene Sandstone: (a) $\theta$ data and fitted model (red 3-parameter and black 4-parameter); (b) measured and predicted $K_r$ using both analytical (solid) and numerical (dashed) solutions.}
\end{figure}

\begin{figure}[h]
 \includegraphics[width=\textwidth]{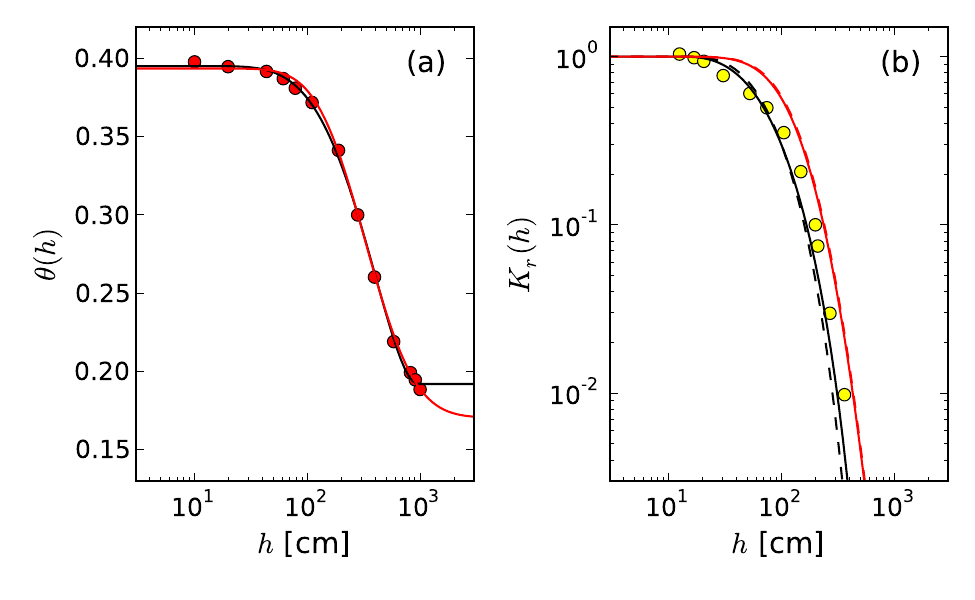}
  \caption{\label{fig:siltloam}  Silt Loam G.E.~3: (a) $\theta$ data and fitted model (red 3-parameter and black 4-parameter); (b) measured and predicted $K_r$ using both analytical (solid) and numerical (dashed) solutions.}
\end{figure}

\begin{figure}[h]
\includegraphics[width=\textwidth]{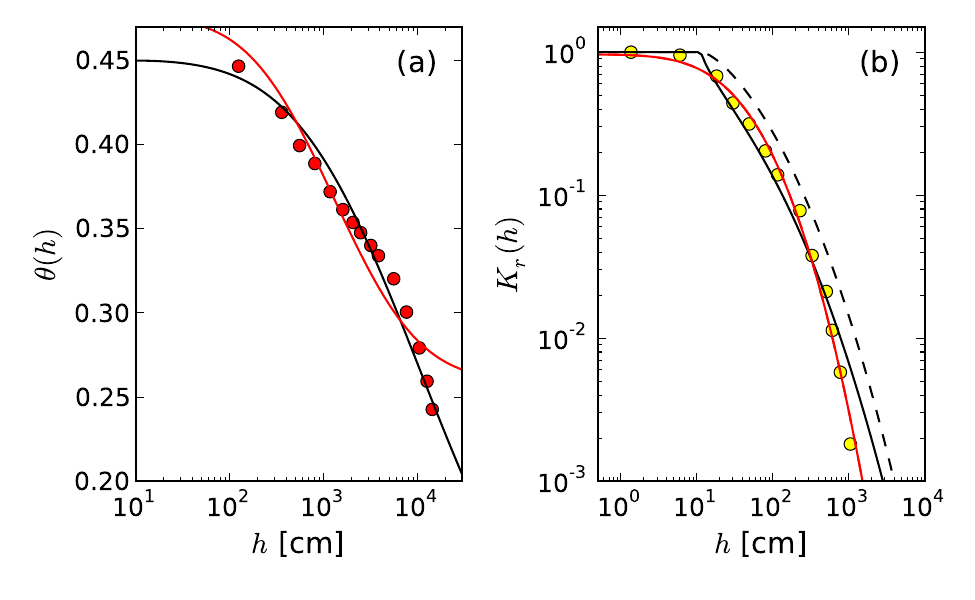}
  \caption{\label{fig:beitnetofa} Beit Netofa Clay: (a) $\theta$ data and fitted model (red 3-parameter and black 4-parameter); (b) measured and fitted $K_r$ using both analytical (solid) and numerical (dashed) solutions for both 3- and 4-parameter models.}
\end{figure}

The three- and four-parameter lognormal models for moisture retention (\ref{eqn:proposed_mrc1}) and (\ref{eqn:proposed_mrc2}) were fitted to experimental data (also used by \citet{vangenuchten1980} and \citet{kosugi1996}) via a quasi-Newton optimization algorithm \citep{zhu1997} from scipy \citep{oliphant2007}. The parameters $r_\mathrm{max}$, $\mu_Z$, $\sigma_Z^2$, $\theta_r$, and $\theta_s$ were estimated for the three-parameter model, while $r_0$ was additional estimated for the four-parameter model.  Model-predicted unsaturated hydraulic conductivity (\ref{eqn:lognK_kosugi}) and (\ref{eqn:lognK_proposed}) were compared to measured values. Figures \ref{fig:hygsanstone} and \ref{fig:siltloam} show data (dots) and best-fit models (lines) for Hygiene sandstone (\citet{mualem1976} soil 4130) and Silt Loam G.E.~3 (\citet{mualem1976} soil 3310). The 3-parameter model is plotted in red and the four-parameter model is plotted in black. Table \ref{tab:estimated_parameters} provides a summary of estimated parameters for the three soils and two models. Figures \ref{fig:hygsanstone}b, \ref{fig:siltloam}b, and \ref{fig:beitnetofa}b show solid curves representing the approximate, but closed-form expression for $K_r$ (\ref{eqn:mualem3}), and dashed curves representing the numerically integrated Mualem relationship (\ref{eqn:mualem}) for comparison. The numerically integrated form takes several orders of magnitude more time to evaluate than the closed-form expression, making its direct use in numerical models impractical. The numerical and analytically derived curves are nearly identical for the sandstone (Figure \ref{fig:hygsanstone}) and silt (Figure \ref{fig:siltloam}). The fits are comparable to those of \cite{kosugi1996} and \cite{vangenuchten1980} for these same soils. 

The three- and four-parameter lognormal models were simultaneously fitted to moisture retention and unsaturated hydraulic conductivity data for Beit Netofa clay (\citet{mualem1976} soil 1006). The four-parameter model (black line in Figure \ref{fig:beitnetofa}) simultaneously fitted both unsaturated conductivity and moisture retention data. The three-parameter model only either fit hydraulic conductivity data (Figure \ref{fig:beitnetofa}) or moisture retention data (not shown -- see examples in \citet{kosugi1996}); we could not get the three-parameter model to fit both data sets simultaneously. Figure \ref{fig:beitnetofa}b shows the difference between the closed-form solutions (solid lines) and the numerically integrated form (dashed lines). The closed-form and numerical solutions for the 3-parameter model are identical (a single red line). The numerically integrated 4-parameter model deviates more significantly from its closed-form counterpart. 

\section{Discussion}
We present a three-parameter approximate closed-form expression for the hydraulic conductivity associated with the three-parameter moisture retention curve of \citet{kosugi1994,kosugi1996}. We use this approach to develop an analogous four-parameter lognormal model, which provides fits to moisture retention data from three soils, similar to the fits with the widely used models of \cite{vangenuchten1980} and \cite{kosugi1996}. Predictions of unsaturated hydraulic conductivity from model fits to moisture retention data are comparable to those of \cite{vangenuchten1980} and \cite{kosugi1996} for Hygiene sandstone and Silt Loam G.E.~3. The four-parameter model yielded a different estimate of rmax than the three-parameter model did (Table \ref{tab:estimated_parameters}), and it predicted the unsaturated permeability curve better (Figure \ref{fig:siltloam}b). Estimating model parameters for Beit Netofa clay using only moisture retention data did not yield good predictions of unsaturated hydraulic conductivity (similar to the models of \cite{vangenuchten1980} and \cite{kosugi1996}). It was essential to use both moisture retention and hydraulic conductivity data to estimate the parameters and improve the fit of the proposed approximate analytical model over the models of \cite{kosugi1994} and \cite{vangenuchten1980}. Moisture retention data alone are not sufficient to estimate all four (or six, when including $\theta_r$ and $\theta_s$) model parameters. If conductivity measurements are available for clays, they should be used with moisture retention data to arrive at a more realistic closed-form model for unsaturated soil moisture retention and hydraulic conductivity.

The difference between the proposed approximate analytical expressions for 3- and 4-parameter lognormal unsaturated conductivity and their much more computationally expensive numerically integrated counterparts is minimal for the two non-clay soils. The models provide useful alternative formulations to the only closed-form lognormal model (with 2 parameters) given by \cite{kosugi1996}. To rigorously verify the validity of the approximate closed-form expressions for a given set of parameters, we suggest comparing them to the results of numerically integrating (\ref{eqn:mualem}), as was done in Figures \ref{fig:hygsanstone}--\ref{fig:beitnetofa}.

One would expect a model with a larger number of adjustable parameters to fit observed data better than similar models with fewer parameters, but improved fit is often at the expense of parameter physical significance. The proposed four-parameter model is more physically plausible model than the simpler two- and three-parameter forms, which make the simplifying assumption of infinite minimum or maximum pore sizes. Models with too few parameters to adequately explain observed data (e.g., the three-parameter model in the case of the Beit Netofa Clay) are structurally deficient; other parameters may take on physically unrealistic values to compensate for the structural deficiency of the model. Likewise, a model with too many parameters will have high uncertainty associated with the unnecessary parameters. For any given soil, the simplest appropriate model should be used. All parameters follow the philosophy \cite{kosugi1994} used in deriving his physically based model: they are related to pore-size distribution statistics and limits, rather than being exponents or powers (i.e., fitting parameters). The modified model of \cite{mualem1976} enables approximate but closed-form expressions for unsaturated hydraulic conductivity for both the three- and four-parameter lognormal pore size distributions. The model derived here is a generalization of the lognormal model of \cite{kosugi1996} for non-zero minimum pore radius and non-infinite maximum pore radius.

\section*{Acknowledgments}
Sandia National Laboratories is a multi-program laboratory managed and operated by Sandia Corporation, a wholly owned subsidiary of Lockheed Martin Corporation, for the U.S. Department of Energy's National Nuclear Security Administration under contract DE-AC04-94AL85000.


\newpage

\begin{table}[ht]
\caption{\label{tab:estimated_parameters}Estimated parameter values for proposed 4-parameter model}
\begin{tabular}{lcccc|cc}
\hline 
Medium & $r_0 (\mathrm{m})$ & $r_\mathrm{max}(\mathrm{m})$ & $\mu_Z$ & $\sigma_Z$ & $\theta_s$ & $\theta_r$\\
\hline 
\multirow{2}{*}{Hygiene Sandstone} & $1.07\times 10^{-4}$ & $2.52\times 10^{-3}$ & $-6.30$ & $0.337$ & $0.250$ & $0.153$\\
                  &                     &  $3.05\times 10^{-3}$ & $-6.30$ & $0.336$ & $0.251$ & $0.147$ \\
\hline
\multirow{2}{*}{Silt Loam G.E.~3} & $1.48\times 10^{-4}$ & $1.27\times 10^{-2}$ & $-7.93$ & $1.12$ & $0.395$ & $0.192$\\
                 &                     & $3.11\times 10^{-2}$ & $-7.69$ & $0.805$ & $0.394$ & $0.171$\\
\hline
\multirow{2}{*}{Beit Netofa Clay} & $4.36\times 10^{-7}$ & $1.32\times 10^{-2}$ & $-11.1$ & $2.33$ & $0.450$ & $0.100$\\
                 &                     & $1.24\times 10^{-3}$ & $-11.0$ & $2.57$ & $0.444$ & $0.119$\\
\hline
\end{tabular} 
\end{table}

\end{document}